\input harvmac

\def\p{\partial}

\def\half{{1\over 2}}

\Title{hep-th/0106184}{\vbox{\centerline{Matrix Model for De Sitter}}}
\vskip20pt

\centerline{Miao Li}
\vskip 10pt
\centerline{\it Institute of Theoretical Physics}
\centerline{\it Academia Sinica}
\centerline{\it Beijing 100080} 
\centerline { and}
\centerline{\it Department of Physics}
\centerline{\it National Taiwan University}
\centerline{\it Taipei 106, Taiwan}
\centerline{\tt mli@phys.ntu.edu.tw}

\bigskip

Based on a heuristic boost argument, we propose that the 4
dimensional de
Sitter space can be described by a spherical Chern-Simons
matrix model near the cosmological horizon, or models
generalizing this simple choice. The dimension
of the Hilbert space is naturally finite. We also make
some comments on possible realization of holography in
this approach, and on possible relation to the conformal
field theory approach. 

\Date{June 2001}

\nref\tb{T. Banks, ``Cosmological Breaking of Supersymmetry?",
hep-th/0007146;
T.Banks~and W. Fischler, ``M-theory observables for cosmological 
space-times",  hep-th/0102077.}
\nref\ew{E. Witten, Talk at Strings' 2001, Tata Institute,
Mumbai, India; ``Quantum gravity in De Sitter space", 
hep-th/0106109.}
\nref\bhm{V. Balasubramanian, P. Horava and D. Minic,
``Deconstructing de Sitter", JHEP 0105 (2001) 043; 
hep-th/0103171.}
\nref\andy{A. Strominger, ``The dS/CFT correspondence", 
hep-th/0106113.}
\nref\desi{J. Maldacena and A. Strominger, ``Statistical entropy
of de Sitter space", JHEP 9802 (1998) 014, gr-qc/9801096;
S. Hawking, J. Maldacena and A. Strominger, ``DeSitter entropy,
quantum entanglement and AdS/CFT", JHEP 0105 (2001) 001, 
hep-th/0002145;
M. Banados, T. Brotz and M. E. Ortiz, ``Quantum
three-dimensional de Sitter space", Phys. Rev. D 59 (1999) 046002,
hep-th/9807216;
W. T. Kim, ``Entropy of $2+1$ dimensional de Sitter space in
terms of brick wall method", Phys. Rev. D 59 (1999) 047503,
hep-th/9810169;
F. L. Lin and Y. S. Wu, ``Near horizon Virasoro symmetry and the
entropy of de Sitter space in any dimension", Phys. Lett. B 453 (1999)
222, hep-th/9901147.}
\nref\ana{A. Volovich, ``Discreteness in de Sitter space and quantization
of Kahler manifolds", hep-th/0101176.}
\nref\ast{S. Perlmutter et al., APJ. 483 (1997); A. G. Riess
et al., Astron. J. 116 (1998) 1009; P. M. Garnavich et al., APJ. 
509 (1998) 74; S. Perlmutter et al., APJ. 517 (1999) 565.}
\nref\boo{P. de Bernardis et al., Nature 404 (2000) 955;
S. Hanany et al., APJ. 545 (2000) L1.}
\nref\sussk{S. Hellerman, N. Kaloper and L. Susskind
``String Theory and Quintessence", hep-th/0104180;
W. Fischler, A. Kashani-Poor, R. McNees and S. Paban, 
``The Acceleration of the Universe, a Challenge for String Theory",
hep-th/0104181.}
\nref\ewr{For a review, E. Witten, 
``The Cosmological Constant From The Viewpoint Of String Theory",
hep-ph/0002297.}
\nref\bou{R. Bousso, ``Positive vacuum energy and the N-bound",
hep-th/0010252; ``Bekenstein bounds in de Sitter and
flat space", hep-th/0012052.}
\nref\bfss{T. Banks, W. Fischler, S.H. Shenker, L. Susskind,
``M Theory As A Matrix Model: A Conjecture", 
Phys.Rev. D55 (1997) 5112, hep-th/9610043.}
\nref\suss{see for example, L. Susskind, ``Strings, Black Holes and
Lorentz Contraction",  Phys.Rev. D49 (1994) 6606,
hep-th/9308139.}
\nref\tpm{``Black holes: The membrane paradigm, edited by 
K. S. Thorne, R. H. Price and D. A. Macdonald, (Yale University
Press, London, 1986).}
\nref\mst{J. McGreevy, L. Susskind and N. Toumbas,
``Invasion of the Giant Gravitons from Anti-de Sitter Space",
JHEP 0006 (2000) 008, hep-th/0003075.}
\nref\hl{P. M. Ho and M. Li, ``Fuzzy Spheres in AdS/CFT Correspondence 
and Holography from Noncommutativity", Nucl.Phys. B596 (2001) 259,
hep-th/0005268.}
\nref\hall{L. Susskind, ``The Quantum Hall Fluid and Non-Commutative 
Chern Simons Theory", hep-th/0101029;
A. P. Polychronakos, ``Quantum Hall states as matrix Chern-Simons theory",
JHEP 0104 (2001) 011, hep-th/0103013;
S. Hellerman, M. Van Raamsdonk, ``Quantum Hall Physics = Noncommutative 
Field Theory", hep-th/0103179.}
\nref\ad{L. F. Abbott and S. Deser, ``Stability of gravity with a 
cosmological constant", Nucl. Phys. B 195 (1982).}
\nref\mbl{G. T. Horowitz, E. J. Martinec,
``Comments on Black Holes in Matrix Theory", 
Phys.Rev. D57 (1998) 4935, hep-th/9710217;
T.Banks, W.Fischler, I.R.Klebanov and L.Susskind,
``Schwarzschild Black Holes in Matrix Theory II", 
JHEP 9801 (1998) 008, hep-th/9711005;
M. Li, `` Matrix Schwarzschild Black Holes in Large N limit",
JHEP 9801 (1998) 009, hep-th/9710226;
M. Li and E. Martinec, ``Probing Matrix Black Holes",
 hep-th/9801070.}
\nref\mli{M. Li, ``Dimensional Reduction via Noncommutative Spacetime: 
Bootstrap and Holography", hep-th/0103107.}

\newsec{Introduction}

The problem of formulating a sensible theory of quantum gravity
on de Sitter space has attracted increasingly more attention 
\refs{\tb-\andy}. Some earlier attempts include \desi.
The cause of these interests seems twofold.
For one thing, string theorists have gained much control in the past 
few years over string theory when there is enough supersymmetry, 
the spacetime background in this case is either Minkowski or 
anti-de Sitter, and there is a proposed nonperturbative formulation
in each case, namely Matrix Theory or Maldacena's CFT description. 
As soon as supersymmetry is completely broken, we have not yet
succeeded in finding a Minkowski background, and most likely,
as proposed in \tb, we will end up with a background with a positive
cosmological constant. Several theoretical difficulties inherent
to such a background have been discussed in \refs{\tb,\ew,\bhm}.
Among these, the problem of formulating observables is particularly
challenging. On the other hand, several observational results
\refs{\ast,\boo} point to the possibility that we are living in
an accelerating universe with considerable amount of dark energy,
This has induced disquieting concerns in the string community
that such reality might contradict our current understanding
of string theory \sussk. 

One encounters two seemingly insurmountable difficulties in trying
to embed de Sitter space in string theory. First, how to generate
a vanishingly small cosmological constant? \ewr. Second, even if this can
be done, how to describe physics in a de Sitter background?
These two problems are likely to be deeply intertwined.
In this note we shall follow the philosophy that we may still construct
a sensible theory given a de Sitter background, temporarily forgetting
the hard problem of generating a small cosmological constant.

One of the first things coming to sight when one tries
to formulate a theory in the de Sitter background is that 
the vacuum structure of de Sitter ought to be more complicated
than that of either Minkowski or anti-de Sitter, since
the entropy of the background as seen by a co-moving
observer is nonvanishing ($>10^{120}$). There is a tiny Hawking
temperature. However, one can not attribute the huge amount of
entropy to thermal fluctuations at the scale of the Hawking temperature,
since on general grounds of thermodynamics the entropy
of this kind is of order 1 only. Also, the thermal fluctuation
of this kind in principle is not observable at all, since the wave
length is the size of the horizon $R\sim H^{-1}$, while a co-moving
observer has to be well-localized. In fact, even the entropy due
to baryonic matter is much smaller than the Bekenstein-Hawking
entropy.

We are thus forced to propose that the entropy, or the large Hilbert
space, is mainly composed of unusual degrees of freedom which
we shall call wee-degrees. Their dynamics must be essentially different
from ordinary particles of massless or massive, yet they are
coupled to ordinary matter. For having a consistent theory,
it is better to postulate that ordinary matter is collective
excitations of wee-degrees.  This requirement is supported by 
the validity of the so-called D-bound \bou, which states that
the entropy of any matter system within the cosmological horizon
plus the Bekenstein Hawking entropy should be no larger than
the Bekenstein-Hawking entropy after the matter system disappears
from the horizon. In terms of our picture, this is just
$S_{matter}+S_{wee}$ increases during the course when the
matter system falls out the horizon and is bounded by $S_{wee}$,
the entropy of the resulting empty de Sitter space inside
the final horizon. A stronger bound, the holographic bound,
also implies that the wee-degrees of freedom should account
for the whole entropy enclosed in the cosmological horizon.

There are at least two criteria for an acceptable candidate theory
of wee-degrees: 1. the theory must account for the large entropy;
2. when a normal matter degree is coupled to the wee-degrees, 
the effective equation of motion reproduced must be identical to
one in de Sitter space. We will take a step toward meeting
the first criterion. In the next section we will present a 
boost argument borrowed from the construction of matrix theory
\bfss, we argue that a matrix model of first order in time derivatives
is a reasonable candidate. We then proceed in sect.3 to propose
a simplest possible
matrix model, the Chern-Simons matrix model on the stretched
horizon, to describe the de Sitter space in 4 dimensions.
De Sitter boost properties and a conformal light-cone frame
are discussed in sect.5. In sect.6 we argue that the light-cone
coordinate is the best choice of time in the matrix model,
and it is related to dilation generator on the infinite past
Euclidean surface. We conclude this paper in sect.7 with some
discussions.

\newsec{Intuitive Considerations}

Suppose we study a system immersed in an environment of constant
energy density $\rho$ in a volume $V$, and this environment is Lorentz 
invariant. 
When the system is at static, the total momentum $P=0$, the total
energy is $M+\rho V$. Boost the system, there will a net
momentum $P$ coming only from the system, the environment does
not contribute. Now there are two possible on-shell conditions,
depending on which view-point we are going to take. The first is
that the system is completely decoupled from the environment,
so the energy of the system and the momentum satisfy
\eqn\onshell{E^2=P^2+m^2.}
We do not get anything interesting from this relation.
Another possibility is to assume that the system in
fact is kind of collective excitation of the environment,
so the total energy and total momentum satisfy a single
on-shell condition
\eqn\onsh{(E+\rho V)^2=(M+\rho V)^2 +P^2.}
If the energy in the environment is much greater than
$M$ and $P$, we obtain, approximately
\eqn\nonre{E=M+{P^2\over 2(PV+M)}+O(P^4),}
in this way, the system behaves as a massive
system with a large mass $\rho V+M\sim \rho V$.
This intuition is quite right, since this implies
that it is extremely hard to excite the environment
whose prototype is the non-dynamic vacuum energy, or
the cosmological constant.

To imitate matrix theory, we go to the limit in which
the longitudinal momentum $P$ is either at the same
scale of $\rho V$ or even much greater than $\rho V$.
Denote $P^2_{eff}=(\rho V)^2 +P^2$, then
\eqn\imf{E+\rho V-P_{eff}={M^2\over 2P_{eff}}
+{\rho VM\over P_{eff}}.}
Since $M\ll \rho V$, the second term dominates
the first term on the R.H.S. of the above equation.
If one names $E+\rho V-P_{eff}=P_-$, it is analogous
to the light-cone energy. 

If something similar to the situation in matrix theory 
holds, then one may identify the first term with
a Hamiltonian similar to that in matrix theory,
that is
\eqn\matr{{M^2\over 2P_{eff}}\sim \tr {1\over 2R}
(\p_t X^i)^2 +V(X),}
where $X^i$ are matrices corresponding to
the transverse coordinates, $R$ is some IR cut-off
in the longitudinal direction, which may be identified
with the horizon size. 
Let us push this analogue a little further, then the
second term, the more important one, is to be
identified with
\eqn\craz{{\rho VM\over P_{eff}}=({\sqrt{2}\rho V
\over \sqrt{P_{eff}}})({M\over \sqrt{2P_{eff}}})
\sim {\sqrt{2}\rho V
\over \sqrt{P_{eff}}}(\tr{1\over 2R}
(\p_t X^i)^2 +V(X))^{{1\over 2}}.}
The square root in the last equality is not an analytic function, so 
itself does not make sense as a candidate for the dominant term
in the Hamiltonian. We however take the square root as 
suggesting that the Hamiltonian should be of first order
in derivatives of $X^i$. In the supersymmetric
matrix theory, the Hamiltonian is the square of a super
charge. This super charge can not be used as a candidate
for our Hamiltonian, since in the first place there 
is no supersymmetry in a de Sitter space, and secondly
the super charge is fermionic so it can not serve as
a Hamiltonian.

In any case, we are tempted by the possibility that in
a certain frame, the Hamiltonian of wee-degrees is that
of matrices and is at most of the first order in time 
derivatives of these matrices.  Suppose the action
principle is applicable here, then the Hamiltonian should
be independent of the time derivatives at all, since
the action is at most of the first order in the time
derivatives. This statement is not modified when we
work in the de Sitter space with judiciously chosen
coordinates, as we shall see in the next sections.

For now let us simply postulate that in a certain frame
the wee-degrees are described by matrices with an action
first order in time derivatives. The general form is
\eqn\acgen{S=\int dt \tr\left (P^i(X)\p_t X^i-V(X)\right).}
The canonical conjugate of $X^i$ is the transpose of
$P^i(X)$, a function of $X$ only. The Hamiltonian is equal
to $\tr V(X)$. Assume for Hermitian matrices, the Hamiltonian
assumes its lowest value for mutually commuting 
matrices, as in matrix theory. Thus in the ground state
only the diagonal entries of $X^i$ are physical degrees
of freedom.

De Sitter space
is invariant under boost, thus it is not possible
to assign a longitudinal momentum to a diagonal
degree of freedom, since unlike in matrix theory in Minkowski
space, we are trying to describe an empty de Sitter
space and possible excitations within. Thus the 
rank of matrices $X^i$ is not to be
interpreted as the total longitudinal momentum or
a certain charge.

We shall argue later that $X^i$ should be treated either
as the transverse coordinates in the steady-state
metric of de Sitter, or the spherical coordinates in
the static metric. In either case there are two coordinates,
and the eigenvalues are subject roughly to the
condition
\eqn\const{(X^1_a)^2+(X^2_a)^2 \le R^2,}
where $X^i_a$ is the $a$-th eigenvalue, and $R$ is the
horizon radius. Let the corresponding eigenvalue of
$P^i$ be $P^i_a$, it is not independent of $X^i_a$
in general. So the phase space of the $a$-th eigenvalues
of $(X^i)$ is two dimensional. To have a compatible
Poisson structure, we require
\eqn\compa{{\p P^1_a\over \p X^2_a}=-{\p P^2_a\over \p X^1_a},}
so the Poisson bracket between $X^1_a$ and $X^2_a$
is
\eqn\pois{\{X^1_a,X^2_a\}=({\p P^1_a\over \p X^2_a})^{-1}.}
The dimension of the Hilbert space
of the $a$-th single eigenvalues is, on the semiclassical level
\eqn\sind{\hbox{dim}H_a ={1\over 2\pi}\int |{\p P^1_a\over 
\p X^2_a}|dX^1_adX^2_a.}
This integral is finite for the integration range
of $X^i_a$ is finite.

The simplest choice of $P^i$ is $l^{-2}\epsilon^{ij}X^j$,
where $l$ is a length scale. This is just the case when
the kinetic term in the action is Chern-Simons like.
To avoid double counting of the phase space, the
kinetic term is actually $\half\int dt\tr P^i\dot{X^i}$.
The dimension of the Hilbert space of the $a$-th eigenvalues
in this case is
\eqn\csdim{\hbox{dim}H_a=h={R^2\over 2\pi l^2}.}

To compute the dimension of the whole Hilbert space,
we now need to assign certain statistics to the eigen-values.
Let $N$ is the rank of matrices $X^i$. If all
eigen-values are distinguishable, the whole Hilbert space
of the lowest energy is
\eqn\infst{H=(H_a)^N.}
If, due to gauge symmetry, the eigenvalues are treated as
bosons, then the whole Hilbert space is
\eqn\bst{H=\hbox{Sym}(H_a)^N,}
where we need to symmetrize the tensor product. In the
first case, the dimension of the total Hilbert space is
\eqn\infdim{\hbox{dim}H=(\hbox{dim}H_a)^N
=\exp \left(N\ln {R^2\over (2\pi l)^2}\right).}
To equal the exponent to the Bekenstein Hawking entropy,
we must have
\eqn\infn{N\sim S\sim {R^2\over l_p^2},}
where we have ignored the factor $\ln [R^2/(2\pi l)^2]$,
since it is not a large number. Even for $l=l_p$,
in reality this factor is about $120\ln 10$.
We conclude that in the ``infinite statistics" case,
the rank of the matrices must be the order of the
entropy \foot{We caution here that sometimes in the
literature this statistics is called Boltzmann statistics,
this however is a misnomer, since the traditional Boltzmann
statistics treats particles as identical, but assumes
that the particle number is much smaller than the
the number of single particle states, so it is
the high temperature limit of both Bose-Einstein statistics 
and Fermi-Dirac statistics.}.

If eigenvalues obey Bose-Einstein statistics, the dimension
of the total Hilbert space is
\eqn\bdim{\hbox{dim}H={(N+h-1)!\over N!(h-1)!}.}
We shall always assume both $N$ and $h$ be
large numbers. Using the Sterling formula,
\eqn\ster{S=\ln\hbox{dim}H=N\ln (1+{h\over N})
+h\ln (1+{N\over h}).}
If $N\ll h$, then the second term is about $N$, so
the first term dominates, this requires that $S\sim
N$. If $N\gg h$, the second term dominates, and
$S\sim h$. In either case, the entropy is at the
same order of the smaller number of $(N,h)$. To us,
the most likely case is $N\sim h\sim S$. For
the Chern-Simons model, we demand $l\sim l_p$,
and $N\sim R^2/l_p^2$. We shall argue in later sections
that this is a reasonable choice.

To conclude, we see that within in the framework
of the first order action, it is possible to account
for the cosmological entropy whether or not the
eigen-values are treated as bosons. In both cases,
the rank $N$ must be not too smaller than the entropy.
It is nonetheless
important that the dimension of the Hilbert space
of the single eigenvalue is finite, otherwise
the dimension of the total Hilbert is infinite.

\newsec{The Spherical Matrix Chern-Simons Model}

De Sitter space of $D$ dimensions can be regarded as
a hyper surface in a $D+1$ dimensional Minkowski
space with the constraint
\eqn\contt{-X_0^2+\sum X_i^2=R^2.}
The De Sitter metric is the induced metric. There
are three well-known coordinate systems. The global
coordinates are
\eqn\glcoor{X_0=R\sinh \tau, \quad X_i=R\cosh\tau x_i,}
where $x_i$ form a $D$ dimensional unit vector thus
parameterize the unit sphere $S^{D-1}$. The metric is written
as
\eqn\glme{ds^2=R^2\left(-d\tau^2 +\cosh^2\tau d\Omega^2_{D-1}
\right).}
Another set of coordinates is
\eqn\stst{\eqalign{t&=R\ln{X_0+X_D\over R}, \cr
x_i&={RX_i\over X_0+X_D}, \quad i<D ,}}
the definition of $t$ demands $X_0+X_D>0$ thus these
coordinates cover only half of the de Sitter space.
The metric in terms of them is the one for the steady-state 
universe
\eqn\stead{ds^2=-dt^2+e^{{2t\over R}}dx_i^2.}
We shall discuss some properties of Killing vectors
of this metric in sect.5.

Finally, the coordinates we will mostly use in this section
are the ones for the static metric
\eqn\statc{t=-{R\over 2}\ln {X_0^2\over (X_0+X_D)^2},
\quad x_i=X_i, i<D.}
The metric reads
\eqn\stam{ds^2=-(1-{r^2\over R^2})dt^2+(1-{r^2\over R^2})^{-1}
dr^2+r^2 d\Omega^2_{D-2}.}

In the following we will focus on the 4 dimensional de Sitter
space, though the geometric discussions apply equally to
other de Sitter spaces. An observer sitting at
$r=0$ is a geodesic observer, this corresponds to an observer
in the steady-state universe at $x_i=0$. It takes an infinite
amount of time for a light signal originating from
$r=0$ to reach the horizon $r=R$, in the static coordinates.
In the steady-state universe, the light reaches horizon in
finite time, this time is the same as recorded by the
observer at the origin, so there seems to be a contradiction.
However, it takes infinite amount of time for this light to return 
to the origin, so the observer sees that the light never
gets to the horizon in finite time of his clock.
The meaning of the horizon therefore is invariant.  

As in many works on black holes \suss, we will here advocate
a stretched horizon picture.  This picture is
perceived by the observer at the origin, and again
this picture is invariant with respect to changing coordinate
system. It is however convenient to work with the static
coordinates. In a normal approach, the stretched horizon
is supposed to be located a proper Planck distance away from
the real horizon. Let $r_0$ be the location of the
stretched horizon, its proper distance to $r=R$ is
\eqn\prodis{d=\int^R_{r_0}dr (1-{r^2\over R^2})^{-1/2}
=R\sin^{-1}\sqrt{1-{r_0^2\over R^2}}\sim 
R\sqrt{1-{r_0^2\over R^2}}.}
If $d\sim l_p$, then the red-shift factor
$$\sqrt{1-{r_0^2\over R^2}}$$
is about $l_p/R$. For a macroscopic de Sitter, this
is a tiny number.

Now for a stationary observer near the horizon,
all outgoing matter will get a huge boost, and the
boost factor is given by $\exp(t/R)$. This can be obtained
by going to the tortoise radial coordinate
\eqn\tort{r^*={R\over 2}\ln {1+r/R\over 1-r/R},}
with a metric
\eqn\torm{ds^2={4e^{2r^*/R}\over (e^{2r^*/R}+1)^2}
(-dt^2+dr^{*2})+\dots.}
The trajectory of a particle is roughly
$t=r^*$, so the red-shift factor is 
$\exp(-r^*/R)=\exp(-t/R)$, a particle of energy
$\epsilon_0$ at the origin now has an energy
$\exp(t/R)\epsilon_0$. If we do not introduce
a cut-off on $t$, we eventually put the particle
in the infinite momentum frame. Thus, our intuitive
consideration in the previous section applies,
and we end up with, if correct, a system described
by a first order action. Notice that the fictitious
observers hovering on the horizon can communicate
with the observer at the origin, so for them there
is a single theory.

In the stretched horizon picture \tpm, the boost is
finite, and the boost factor is $R/l_p$. A
wave length $1/R$ is boosted to a wave length
$1/l_p$, a Planckian particle. It is therefore
natural to suppose that the relevant length
scale in the Chern-Simons type action is set
by $l_p$, as we already did in the last section.
In this case, our intuitive boost argument is
to be taken merely as an argument, since 
the boost can not be made infinite. Locally,
we now have a flat Chern-Simons matrix model,
globally, the Chern-Simons model is defined
on a sphere of radius $R$. 

It is tricky to
write down a matrix model with matrices assuming
values on a 2 sphere. We by-pass this problem
by using a specific coordinate system covering
the sphere. Choose the sterographic projection
coordinates $y^i$ on the unit sphere with
a metric
\eqn\stem{ds^2={4(dy^i)^2\over (1+y^2)^2},}
with the symplectic form
\eqn\symf{{4dy^1\wedge dy^2\over (1+y^2)^2}.}
Integrating over the sphere we obtain the area
of unit sphere $4\pi$.
For a pair of given eigen-values, we expect
the Chern-Simons to yield a coupling
\eqn\cscoup{n\int dt f(y^2)\epsilon^{ij}y^i\dot{y}^j,}
where $n\sim R^2/l_p^2$, the factor $R^2$ 
sets in since we are working with a sphere of
radius $R$, and the factor $l_p$, as we just
argued, is due to the fact that we are talking
about Planckian particles.
The action \cscoup\ corresponds a symplectic form
\eqn\resym{n(2y^2f'+2f)dy^1\wedge dy^2.}
Apart from the coefficient $n$, we expect
the above to agree with \symf, so $f$ is solved
to be
\eqn\fsol{f(y^2)=-{2\over y^2(1+y^2)}.}

Semi-classically, the quantization of the
Chern-Simons coupling \cscoup\ results in
a Hilbert space of dimension 
\eqn\sedim{{1\over 2\pi}n\int {4d^2y\over
(1+y^2)^2}=2n,}
so $n$ is either an integer or half integer
at the semi-classical level. The actual value
of $n$ is a little more complicated as we shall
see shortly. 

We need now to promote $y^i$ to
Hermitian matrices $Y^i$, and to postulate
a Chern Simons matrix action
\eqn\csact{S=\int dt\left(-2n\tr {1\over Y^2(1+
Y^2)}\epsilon^{ij}Y^i\dot{Y}^j -V(Y)\right).}
As usual in matrix models defined on a curved space,
there is an ordering ambiguity in the above action.
We postpone discussing this issue to a later occasion.
The potential term $V(Y)$ can not be specified by
our limited knowledge of the microscopic theory.
It is supposed to be rotationally invariant, and
assume minimum when the two matrices $Y^i$ commute.
One can also contemplate the possibility of adding
more matrices such as some fermionic matrices to
the above action, to get a realistic model. We will
not discuss this possibility in this paper. 

It remains to discuss the exact quantization procedure
and construct the Hilbert space for single eigen-values. 
To do so, it is convenient to switch to a more physical
coordinate system. One choice is the Cartesian coordinates
for the sphere $x^i$, $i=1,2,3$, $\sum (x^i)^2=R^2$. The
relation to the sterographic projection coordinates is
\eqn\rels{x^i={2R\over 1+y^2}y^i, \quad i=1,2.}
The corresponding symplectic form
\eqn\csymf{{n\over R\sqrt{R^2-x^2}}dx^1\wedge dx^2,}
where $x^2=(x^1)^2+(x^2)^2$. The Chern-Simons
coupling is
\eqn\pcs{-{n\over R}\int dt {\sqrt{R^2-x^2}\over 
x^2}\epsilon^{ij}x^i\dot{x}^j.}
This is the action of a unit charged particle
moving in a monopole background with a
magnetic charge $n$. The above action can
be generalized to a nonrelativistic charged particle
moving in the full 3 dimensional space
\eqn\cmon{S={m\over 2}\int \dot{x}^2dt
+\int A_i\dot{x}^idt,}
where the second term is to be identified with
\pcs. It is well-known that the angular momentum
of the system is corrected by a magnetic term
\eqn\angm{M^i=\epsilon^{ijk}(mx^j\dot{x}^k
-mx^k\dot{x}^j)-{nx^i\over r}.}
Upon quantizing the action \cmon, one gets nontrivial
commutators among $\dot{x}^i$, while the commutators
among $x^i$ vanish. For instance
\eqn\noncomm{[m\dot{x}^i,m\dot{x}^j]=i\epsilon^{ijk}
B_k,}
where $B$ is the magnetic field generated by
the monopole. One can check that the angular
momentum components satisfy the usual $su(2)$
algebra
\eqn\sut{[M^i,M^j]=i\epsilon^{ijk}M^k.}

Now taking the limit $m\rightarrow 0$, the
first kinetic term in \cmon\ drops out, thus
the canonical momentum is not independent of
$x^i$. The first term in \angm\ also drops
out, the net angular momentum has a contribution
solely from the magnetic field.
In this limit, the quantization can be
carried out by Dirac brackets. We shall not
do this, we shall just put the system on a sphere
of radius $R$ and quantize the action \pcs.
With the symplectic form \csymf, we find
\eqn\canc{[x^1,x^2]=-i{R\sqrt{R^2-x^2}\over n}.}
It is straightforward to check that, to the
first order, the commutation relations
\sut\ still hold, with 
\eqn\siman{M^i=-{nx^i\over R}.}
In fact, to make the quantization procedure
rigorous, we simply assume the above relations.
Thus $x^i$ up to a constant form a $su(2)$
algebra. The Hilbert space is specified by
picking an irreducible representation of spin
$J$. Since the Casimir is 
$$\sum_i (M^i)^2=J(J+1)$$
we have
\eqn\sphcon{\sum_i (x^i)^2={J(J+1)\over n^2}R^2.}
Now $x^i$ represent a fuzzy sphere. If we demand
the radius of this fuzzy sphere to be exactly
$R$, then
\eqn\quanc{n^2=J(J+1).}
The dimension of the Hilbert space is $2J+1$.
In the large $J$ limit, $n\sim J$, so the dimension
of the Hilbert space is approximately $2n$,
a result agreeing with the previous semi-classical
analysis. The horizon as a fuzzy sphere was discussed
previously in \ana.

It should be noted that we are not entitled to assume
the radius of the fuzzy sphere to be $R$, this
is because we are talking about the stretched horizon
whose radius is not yet determined. This question is
equivalent to determining $n$, and further microscopic
knowledge is required in order to do so.

It is clear that what we have from the simple Chern-Simons
model is giant gravitons. The cause of this phenomenon
in our opinion is the vacuum energy whose nature is
still mysterious, while giant gravitons in the AdS/CFT
correspondence arise due to the R-R flux background \mst.
Consequently, the mechanism to obtain the fuzzy sphere
is different from the one in that case \hl.

As we shall see in sect.5 and sect.6, it is even better to
define Chern-Simons matrix model using a light-cone time.
Moreover, the sterographic coordinates will have a direct
physical meaning. In this language, it is possible to 
relate our approach to the approach in \andy.

We have not gauged our matrix model. It is possible
to introduce a gauge matrix $A_0$, and turn the time
derivative in the Chern-Simons action \csact\ into
a covariant derivative. This model becomes closer to
the planar matrix model of \hall. Of course there is
important difference: We have postulated that the
potential $V(Y)$ is so chosen to force matrices $Y^i$
become commuting for lowest energy.

According to discussions in the next section, it appears
necessary to introduce fermionic matrices and perhaps
other bosonic matrices. Once fermionic matrices are
present, it is possible to introduce supersymmetry.
SUSY is likely broken for most of the ground states,
since in these states, the eigenvalues of
$Y^i$ are not homogeneously distributed. It is therefore
very interesting to determine the SUSY breaking
scale dynamically. We will come
back to this important issue in future.

\newsec{The Membrane Paradigm and Holography}

The matrix model proposed in the last section can
be regarded as a concrete realization of the stretched
horizon. As in the membrane paradigm \tpm, we should
be able to assign various physical properties to the
stretched membrane. For instance, consider that there is
an electromagnetic field in the bulk enclosed by the horizon,
the bulk electrodynamics should be completely specified
by the electromagnetic properties of the stretched
horizon, the conductivity of the horizon
thus Ohm's law is a simple example.

More precisely, the boundary conditions for the electric
field and the magnetic field on the stretched horizon
are determined by the charge density and the current
density on the stretched horizon, for instance
\eqn\ohm{E_\perp =4\pi\sigma, \quad B_\parallel 
=4\pi (j\times n),}
where $\sigma$ is the charge density on the sphere, and
$j$ is the current density, $n$ is the unit vector
normal to the sphere.
 
It should be possible to define charge and current density
in the Chern-Simons matrix model if one assigns
specific electro-magnetic properties to the excitations
in the model. One possibility is to identify our
model with some version of quantum Hall system in which
all these quantities are clearly defined. In fact
recently some planar matrix models are proposed to
describe the real quantum Hall system \hall.
One even can go a step further to incorporate
the standard model fields in the bulk, by adding
more matrices to the model to construct corresponding
sources for these fields. However, it seems premature
to construct even the charge density for the time being,
since this is possible only when more details of the
matrix model are given. 

It is also important to understand how holography is
realized in de Sitter space. Following the philosophy
of AdS/CFT correspondence, one need to spell out
a dictionary between operators in the matrix models
and bulk modes. It seems straightforward to do this
once more details are known about the matrix model.
For example, to operators of the type
\eqn\strace{\tr Y^{i_1}\dots Y^{i_n},}
one may associate  spherical harmonic modes of a 
scalar field in the bulk. Apparently, in reality
there are many fields in the bulk, we need to introduce
more matrices in the matrix model in order to account
for all of them along this line.

Whether it is really possible to generalize the AdS/CFT
correspondence to the present context remains to be
seen. An apparent objection to this naive generalization
is that unlike there is no
scaling argument available. One potential argument replacing
the scaling argument is the infinite boost close to
the horizon. Response of a given bulk mode under the
boost may be translated to a certain property of the 
operator in the matrix model. We shall see in sect.6
that matrix model can be built in another set of coordinates,
the conformal light-cone coordinates, one may
find out scaling argument in that context, for instance,
the evolution with the light-cone time is related to
dilation on the infinite past Euclidean space.
Thus, the scaling dimension can be identified with the
eigen-value of the matrix Hamiltonian $H(Y)=V(Y)$:
\eqn\hale{[H(Y), {\cal O}]=h{\cal O},}
now the Hamiltonian is taken to be the one evolving
everything along the light-cone time, for more
details, see sect.6.

\newsec{Some Observations on Boost Transformations}

Given the success of matrix theory in Minkowski space,
one wonders whether it is possible to go to an 
infinite momentum frame in de Sitter space
and to imitate matrix theory
to get a smaller symmetry and write down a matrix
model. Although the observations here are somewhat
interesting, we are led to negative results.

We will be working with the steady-state metric
\eqn\stst{ds^2=-dt^2+e^{{2t\over R}}dx_i^2.}
The apparent Killing vectors are those corresponding
to translations in spatial directions and a shift
in time together with rescaling of spatial coordinates
\eqn\killing{\eqalign{P_i&=\p_i,\cr
H&=\p_t-{x^i\over R}\p_i.}}
The conserved quantity associated with $H$
is the AD energy \ad.
For a massive particle of the action
\eqn\massa{S=-m\int dt\left(1-e^{2t/R}(\dot{x}^i)^2
\right)^{1/2},}
the AD energy is given by
\eqn\ade{E=m\left(1-e^{2t/R}(\dot{x}^i)^2\right)^{-1/2}
[1+{x^i\dot{x}^i\over R}e^{2t/R}]
=(m^2+e^{-2t/R}p^2)^{1/2}+{x^i\over R}p^i.}
If restricted inside the horizon 
$$e^{t/R}|x|\le R$$
the energy \ade\ is always positive.

Although it is possible to use the energy \ade\
to generate a flow, it is easy to see that the
usual Hamilton-Jacobi equations are not valid,
so one can not use \ade\ as a Hamiltonian in the usual
sense. Moreover, $H$ does not commute with
$P_i$.

The remaining Killing vectors are rotations in
space and boost transformations. To construct
a boost in a direction, say $z=x^{D-1}$, we
go the the light-cone frame in which
\eqn\lighc{ds^2={4R^2\over (x^++x^-)^2}
\left(-dx^+dx^-+dx_{\perp}^2\right),}
where the light-cone coordinates are
\eqn\lcoord{x^\pm= -Re^{-t/R}\pm z,}
and the remaining coordinates $x_{\perp}$ are
transverse directions. The metric
\lighc\ is conformal to the Minkowski metric,
and the horizon for an observer at
$x^i=0$ is given by
\eqn\horz{x^+x^--x_{\perp}^2=0.}
The inside of the horizon is specified by
the conditions $x^+<0$, $x^-<0$ together
with $x^+x^--x_{\perp}^2>0$.

The boost transformation generalizing
the flat space boost $x^\pm\rightarrow
\exp(\pm \beta)x^\pm$ is rather involved,
since the transformation acts on the transverse
direction too. For $dS_2$, it is relatively
easy to find out the boost transformation 
which is
\eqn\dstw{x^\pm ={\cosh{\beta\over 2} y^\pm \pm
\sinh{\beta\over 2} R\over \cosh{\beta\over 2}
\pm \sinh{\beta\over 2}y^\pm/R}.}
For small $\beta$, the above transformation
reduces to the one in the flat space, where
for small $t$, $x^\pm$ goes over to 
$t\pm z-R$. For large $\beta$, the nonlinear
effect becomes important, and both $x^\pm$
goes over to constants $\pm R$. This is
easy to understand for $x^-$, since 
$t-z$ shrinks to zero for large $\beta$,
so $x^-$ tends to $-R$. This limit is unfamiliar
for $x^+$, since in the flat case one expects
$x^+$ blow up instead of going over to a 
constant. 
The strange limit of the large boosts tells
us that it is impossible to go to an infinite
momentum frame and meanwhile to restrict everything
within the horizon, because $x^+\rightarrow
R$ and is no longer negative. 

For higher
dimensional de Sitter space, the boost transformation
is much more involved:
\eqn\boostt{\eqalign{x^\pm &={\cosh{\beta\over 2} y^\pm \pm
\sinh{\beta\over 2} R\pm \sinh{\beta\over 2} (fR)^{-1}
y_{\perp}^2\over \cosh{\beta\over 2}
\pm \sinh{\beta\over 2}y^\pm/R},\cr
x_{\perp}&=f^{-1}y_{\perp},}}
where
\eqn\factf{f=(\cosh{\beta\over 2}+\sinh{\beta\over 2}
y^+/R)(\cosh{\beta\over 2}-\sinh{\beta\over 2}y^-/R)
+{1\over 2R^2}(\cosh\beta-1)y_{\perp}^2.}
Although the analysis is complicated, the
conclusion drawn in the previous paragraph still holds,
namely it is impossible to go to the infinite 
momentum frame again.

Although our results are rather undesired, we can still
ask whether there are corresponding longitudinal momenta
which are conserved and scale under a boost just as
in a flat space. The answer to this question is positive.
Take $dS_2$ as an example. The Killing vector associated
to boosting in the $z$ direction can be obtained from
\dstw\ by expanding in $\beta$ to the first order,
\eqn\booki{K={R\over 2}[1-({x^+\over R})^2]\p_+
-{R\over 2}[1-({x^-\over 2})^2]\p_-.}
The algebra formed by this vector with the
two Killing vector in \killing\ is
\eqn\dstk{\eqalign{[E,P]&={1\over R}P,\quad
[P,K]=E,\cr
[E,K]&=P-{1\over R}K.}}
This is of course the de Sitter algebra $so(2,1)
=sl(2,R)$.
Now the conventional longitudinal momenta
$P^\pm =E\pm P$ are no longer eigenstates
of $K$, rather, the following modified longitudinal
momenta are
\eqn\longi{P^\pm=E\pm P\mp {1\over R}K,}
it is rather interesting to see that these conserved
quantities involve the boost generator $K$. Under
a finite boost of rapidity $\beta$, we have
\eqn\longb{P^\pm \rightarrow e^{\pm\beta}P^\pm.}

One might wonder whether one can use the above
boost property to simplify the kinetics of a particle.
Again the answer is negative, since the on-shell
condition in terms of $P^\pm$ is complicated, this
is because, in terms of the conserved momentum $P$
and energy $E$, the boost generator is not
a simple function and involves coordinates directly.

Although the usual infinite momentum frame is 
impossible, it is still possible to implement the
idea of the stretched horizon in the coordinate
system \lighc. Take $x^+$ as the time variable. 
For fixed $x^+$, the horizon is located at
\eqn\loch{x^-={x_{\perp}^2\over x^+}.}
The metric on the horizon is
\eqn\hmetr{ds^2_{hor}={4R^2(x^+)^2\over
((x^+)^2+x_{\perp}^2)^2}dx_{\perp}^2.}
This is the metric on a sphere of radius $R$,
since the dependence on $x^+$ in the metric
can be rescaled away. Now the coordinates
$x_{\perp}$ provide a physical way to realize
the sterographic projection coordinates we introduced
in sect.3 purely for convenience. Thus it may be
more natural to study the Chern-Simons matrix
model in which $x^+$ is taken as time, and the
matrices assume values in the transverse space.
It is even possible to make our intuitive argument
for the Chern-Simons matrix model more concrete
and appealing in the light-cone coordinates.
We shall argue in the next section that
it is better to take $\ln x^+$ as time in a theory
on the horizon.

\newsec{Relation to Strominger's Approach}

Strominger very recently proposed a dS/CFT 
correspondence \andy. In his approach, there is
Euclidean CFT living on the past sphere 
at $\tau=-\infty$ in the global coordinates
\glme. The microscopic details of the CFT can not
be specified except its central charge. 
The CFT is nonunitary, there are complex conformal
weights. 

In case of $dS_3$, the CFT is two dimensional.
It is suggested that states in the Hilbert space
are given in the radial quantization. 
Interestingly, it is shown in \andy\ that one can
identify the AD Killing vector $H$ in \killing\
with the radial generator $L_0+\bar{L}_0$.
Thus, in order to establish any connection between
a horizon theory such as the spherical Chern-Simons
matrix model in sect.3 with the Euclidean conformal
theory at the infinite past, it is important to
identify this Killing vector in the theory on
the horizon.

The AD Killing vector, up to a constant, can be written
in the light-cone coordinates as
\eqn\adh{H=-(x^+\p_++x^-\p_-+x^i\p_i),}
where we omitted the subscript $\perp$ for the 
transverse coordinates. Now we ask, taking $x^+$ as
our time variable, how the AD Killing vector acts
on a function on the horizon. Substitute the constraint
\loch\ into the function $f(x^+,x^-,x^i)$, we obtain
a function 
\eqn\ne{\tilde{f}(x^+,x^i)=f(x^+,{x^2\over x^+},x^i).}
We find that the operators acting on
the function $\tilde{f}$ are related to operators
acting on $f$ in the following way
\eqn\relo{\eqalign{x^+\p_+ &\rightarrow 
x^+(\p_++{x^2\over (x^+)^2}\p_-),\cr
x^i\p_i&\rightarrow x^i(\p_i-{2x^i\over x^+}\p_-).}}
Using these relations we see that when acting
function $\tilde{f}$, the AD Killing vector reduces
to
\eqn\had{H=-(x^+\p_++x^i\p_i).}
This reduction is to be expected, since the
Killing vector in \adh\ is the dilation operator
for all coordinates including $x^-$. Now the
constraint \loch\ respects the dilation, so
when restricted to the horizon, the dilation
operator should be \had.

This AD Killing vector on the horizon can be 
further simplified by noticing that the metric 
restricted on the horizon depends on $x^+$, as
in \hmetr. Performing the rescaling
$x^i\rightarrow x^ix^+$, the metric becomes
\eqn\nhm{ds_{hor}^2={4R^2\over (1+x^2)^2}dx^2.}
Namely, $x^i$ are just the sterographic coordinates
$y^i$ introduced in sect.3. By a similar argument
as the above, we find that now in terms of $x^+$ and
the sterographic coordinates $x^i$, the AD Killing
vector is simply
\eqn\hads{H=-x^+\p_+.}
Now it is clear that, given the fact that $H$ should
be a conserved quantity, we should take
$-\ln (-x^+)$ as the time for a theory on the horizon.
We use $-x^+$ in the logarithmic since inside the
horizon $x^+<0$. The light-cone time $\tau=-\ln(-x^+)$
increases with $x^+$ and tends to $\infty$ as 
$x^+\rightarrow 0$, its range is $(-\infty,\infty)$.

So it is natural to build up the matrix Chern-Simons
model introduced in sect.3 directly on the horizon
in the light-cone coordinates. The matrices $Y^i$ just
correspond to the transverse coordinates discussed in
the previous section and the present section. Furthermore,
the matrix Hamiltonian is identified with the AD
Killing vector on the horizon, thus should be identified with
the radial generator in the Euclidean conformal theory
of Strominger.

Just as Strominger, here we can not say much about
microscopic details of the Hamiltonian in the Chern-Simons
matrix model, which is simply given by the matrix
potential $V(Y)$ in \csact. Although it is just a function
of matrices $Y^i$, its dynamic content is nontrivial since
entries of matrices $Y^i$ are noncommutative.

\newsec{Discussions}

Quantum gravity on de Sitter spaces is not expected to be found in
any time soon, not only because there are conceptual challenges,
but also because if the fate of our universe is such a space,
there are many realistic questions to answer, such as
supersymmetry breaking and fitting the standard model
parameters.  Nonetheless these universes provide an excellent
place to test our ideas about string/M theory and will
help in the long term search for fundamental principles
and underlying degrees of freedom. 

We present one possible holographic theory for 4 dimensional
de Sitter space. According to our experience with matrix
theory, the matrix model is highly dimension-dependent,
so we do not attempt to say anything about de Sitter spaces
in other dimensions. In our opinion, even if our model
will ultimately be proven to be on a wrong track, it is
useful for stimulating other ideas. Further, as we
tried in sect.6, it will be useful to compare our attempt
to other attempts now available or to appear in future.

It is interesting to compare the matrix model of the de Sitter
space with the matrix description of Schwarzschild black holes
in a flat space \mbl. Both descriptions are a version of
the stretched horizon picture. Nevertheless there are
fundamental differences. The first obvious distinction is
that the de Sitter matrix model describes the interior physics
viewed by an observer inside the horizon, while the matrix
black hole describes the black hole as seen by an outside
observer, presumably encoding information about the whole
black hole including its interior, although one is not
supposed to talk about the causally disconnected interior
for the outside observer. Secondly, the de Sitter matrix
model is the whole theory of the de Sitter space, that is, 
its Hilbert space seen by one observer is the whole Hilbert
space of the de Sitter space, according to the cosmological
complementarity principle. The matrix black hole only
describes a subset of excited states within matrix theory,
and itself can not be viewed as a complete, unitary theory.
In fact the black hole states are only metastable. 
The most conspicuous difference is that the de Sitter
space matrix model should not only describe the pure de Sitter
space, but also other states which are asymptotically
de Sitter. Thus, it should also describe a state where there
are a cosmic horizon and a black hole horizon, although
at present it is not clear how to construct such a
state in the matrix model.

The cosmological initial conditions can in principle be discussed
in our matrix model. When nonabelian matrices are sufficiently
excited, the dynamics does not necessarily describe physics
only near the cosmological horizon, since there is no clear
geometric picture of horizon in this situation.

A difficult question is how to realize cosmological 
complementarity principle, namely how to relate the
Hilbert for one observer to the one for another observer.
Perhaps a bulk formulation is needed to order to
establish a bridge. It is not a far-off speculation
that the bulk theory is a theory realizing spacetime
noncommutativity and holography along the line of
\mli.

One may speculate a lot more about things related to our approach, 
but it is safe to leave many interesting issues for the future.

Acknowledgments. 
This work was supported by a grant of NSC, and by a 
``Hundred People Project'' grant of Academia Sinica and an 
outstanding young investigator award of NSF of China. 
I thank Y. H. Gao, P. M. Ho and Y. S. Wu for helpful
discussions.

\vfill
\eject

\listrefs
\end